\documentclass[runningheads]{svmult}

\usepackage{makeidx}   
\usepackage{graphicx}  
\usepackage{subeqnar}  
\usepackage{multicol}  
\usepackage{physprbb}  
\makeindex             

\newcommand{\ebv}{{\ensuremath{E(\mathrm{B}-\mathrm{V})}}}

\newcommand{\lte}{\ensuremath{\log(\mathrm{T}_{eff})}}
\newcommand{\ha}{\ensuremath{\mathrm{H}\alpha}}
\newcommand{\ie}{\emph{i.e.}\ }
\newcommand{\arcsec}{\ensuremath{^{\prime\prime}}}

\begin{document}
\title*{TTauri Stars in the Large Magellanic Cloud:\\ a combined HST and
VLT effort}
\toctitle{TTauri Stars in the Large Magellanic Cloud:
\protect\newline a combined HST and VLT effort}
\titlerunning{TTauri stars in the Large Magellanic Cloud}
\author{Martino Romaniello\inst{1}
\and Nino Panagia\inst{2}
\and Salvatore Scuderi\inst{3}
\and Roberto Gilmozzi\inst{4}
\and Eline Tolstoy\inst{5}
\and Fabio Favata\inst{6}
\and Robert P. Kirshner\inst{7}}
\authorrunning{Martino Romaniello et al}
\institute{European Southern Observatory -- Garching bei M\"unchen (Germany)
\and Space Telescope Science Institute -- Baltimore (USA)
\and Osservatorio Astrofisico -- Catania (Italy)
\and European Southern Observatory -- Paranal Observatory (Chile)
\and Gemini Support Group -- Oxford (United Kingdom)
\and ESTEC -- Noordwijk (The Netherlands)
\and Harvard-Smithsonian Center for Astrophysics -- Boston (USA)}

\maketitle

\begin{abstract}
The combination of the unprecedented spatial resolution attainable with
WFPC2 on board HST and of the large collecting area of the VLT makes it
possible to study in detail the low mass pre-Main Sequence stars in
galaxies other than our own.
Here we present the results of our studies of two star forming environments
in our closest galactic neighbor, the Large Magellanic Cloud: the region
around Supernova~1987A and the double cluster NGC~1850.
\end{abstract}

\section{Stellar Populations in the Large Magellanic Cloud}
When it comes to studying stars in galaxies other than our own Milky
Way the Large Magellanic Cloud (LMC) is, for several reasons, an obvious
starting point:

\begin{itemize}
\item With a distance of $52\pm1$~kpc \cite{rom00} it is the closest galaxy
  we can look at from the outside and it is fairly easy to reach down
  to stars of 1~M$_\odot$, corresponding to $m_V\simeq24$, or less.
\item All of the stars are at one and the same distance.
\item Our view is not severely obstructed by Galactic extinction:
  \ebv$_\mathrm{Galaxy}=0.05$ \cite{schw91}.
\item The stars in the LMC span a wide range of ages and physical conditions
  from Globular Cluster-like to star forming environments.
\item Low metallicity: Z$\simeq$Z$_\odot/3$ corresponds to the mean
  metallicity of ISM at $z\simeq 1.3$ \cite{pei99} at which
  the overall star formation rate is highest \cite{madau96}.
\end{itemize}

\begin{figure}[!ht]
\begin{center}
\includegraphics[width=0.7\textwidth]{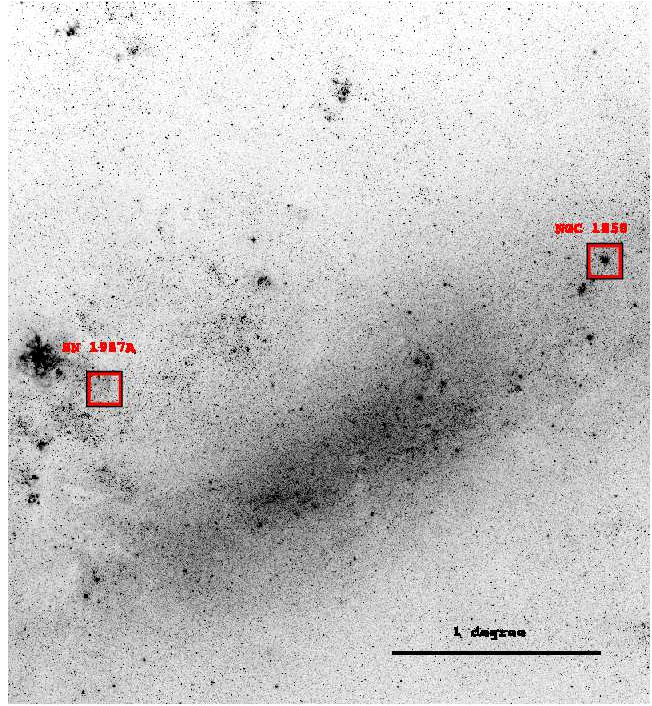}
\end{center}
\caption[]{The location of SN~1987A and NGC~1850 in the LMC on a DSS image
of the galaxy. The scale is shown by the horizontal bar.}
\label{fig:lmc-dss}
\end{figure}

The location in the LMC of the two regions we have studied, the surroundings
of SN~1987A and the double cluster NGC~1850, are shown in
Fig.~\ref{fig:lmc-dss} on a Digitized Sky Survey image of the galaxy.

\section{HST-WFPC2 Imaging}
\subsection{The Region of Supernova~1987A}
The first star forming region we have considered in our search for low mass
pre-Main Sequence (TTauri) stars is the one around SN~1987A. The pre-Supernova
evolution of its progenitor, Sk~-69~202 \cite{sand69}, is estimated to have
lasted 10-12~Myr \cite{vandyk98} and one can expect to find a similarly
young population, born together with it.

\begin{figure}[!ht]
\begin{center}
\includegraphics[width=0.45\textwidth]{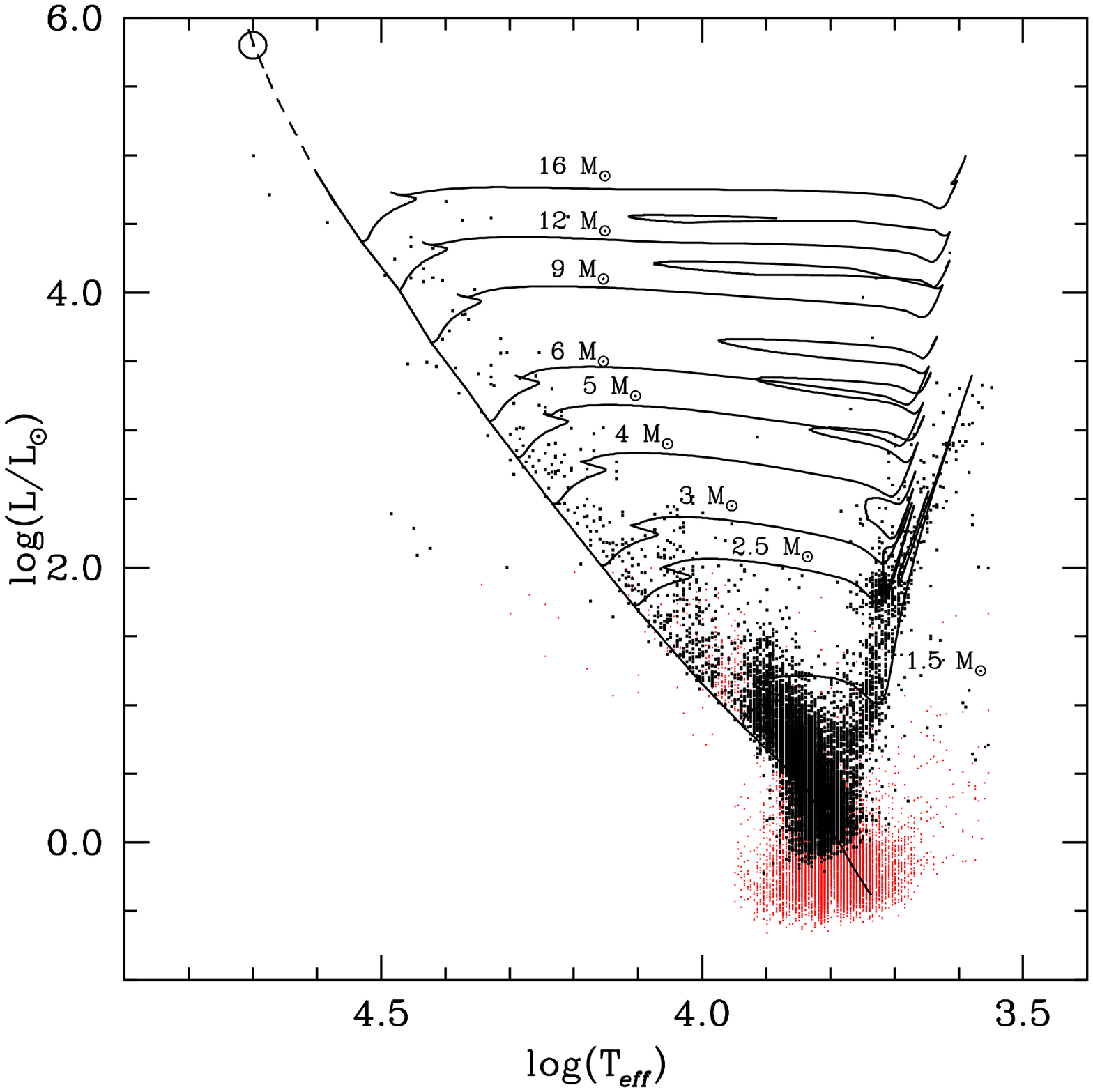}
\includegraphics[width=0.45\textwidth]{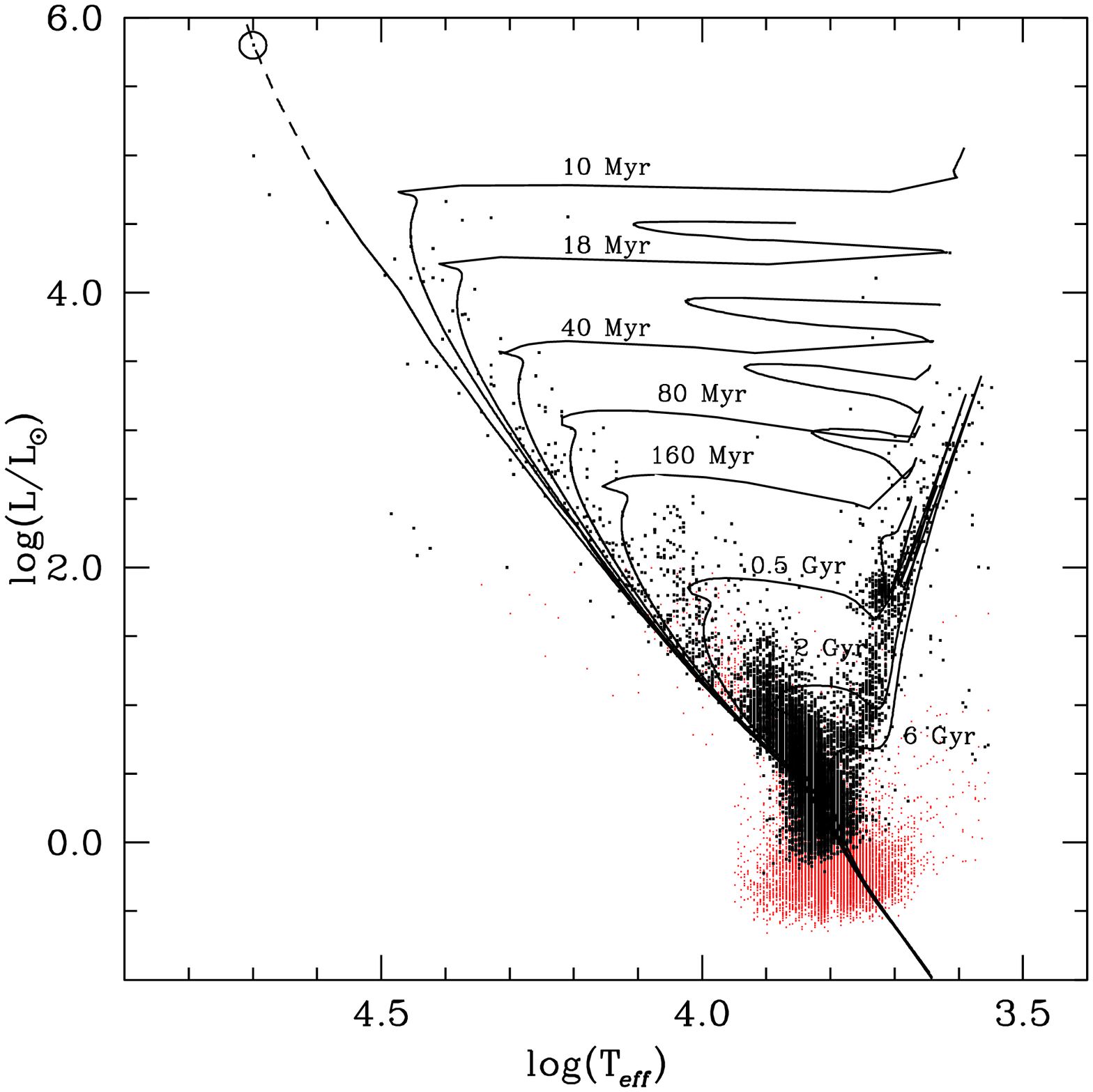}
\end{center}
\caption[]{HR diagram for the stars in the field of SN~1987A. Black
dots are stars with $\delta\lte<0.05$, while the circle highlights
the most massive star in the field. Evolutionary tracks (left panel) and
isochrones (right panel) by \cite{bc93} and \cite{cass94} are overplotted
to the data. The dashed line is the upper Main Sequence as computed by
\cite{sch93}.}
\label{fig:s87a_hr}
\end{figure}

From 1994 the region was imaged almost every year with the WFPC2 as a
part of the long term {\bf S}upernova {\bf IN}tensive {\bf S}tudy led
by Bob Kirshner. This resulted in the coverage in 6 wide bands,
from 2500 to 8500~\AA, plus OIII 5007~\AA, \ha\ and NII 6548~\AA, of a
circular region with a radius of 30~pc centered on the Supernova remnant.
The HR diagram for the 21,955 stars we have identified in our multiband
WFPC2 frames and for which we have derived accurate luminosities and
temperatures with a new technique based on photometry alone \cite{rom01} is
shown in Fig.~\ref{fig:s87a_hr}.

As shown in Fig.~\ref{fig:s87a_hr} there are stars of very different ages,
ranging from a few million to several billions years. In particular, the
location of the most massive stars in the field, except for the one highlighted
in the circle, is consistent with them being coeval to the progenitor of
SN~1987A which, indeed, was not born in isolation, but, rather, in a loose
cluster \cite{pan00}. These massive ($M\simeq 12M_\odot$), bright
($\log(L/L_\odot)\simeq4.5$) stars are easy to identify even in a region
of complex star formation such as this one. Unfortunately, though, this is
not the case for the corresponding low mass population. The expected
location in the HR diagram for these stars ($\lte\simeq3.8$,
$\log(L/L_\odot)\simeq0.2$), which are still contracting towards the Main
Sequence, overlaps with the one of the (much more numerous) field sub-giants
that, with a similar mass of a few solar masses, but an age of several
billion years, have just left it.

The fundamental issue, which will present itself each and every time
a star forming region is projected onto a much older population, then,
is to find a way to identify the TTauri stars and disentangle them from the
sub-giants. Luckily, the so-called Classical TTauri stars, which are
thought to have a disk around them, have at least two clear, distinctive
and correlated characteristics in the optical: a U-band excess when compared
to a photosphere of an evolved star of the same spectral type (see, for
example, \cite{gul98}) and an \ha\ emission which can amount to several tens
of Angstroms (see, for example, \cite{edw94}).

Using these diagnostic tools we identified 850 TTauri candidates with U-band
excess and 488 candidates with \ha\ emission. The vast majority of these
latter ones also show an excess in the U. Let us state here very clearly
that \emph{both the criteria mentioned above will for sure underestimate the
real number of TTauri stars}. On the one side, the detection level will
in both cases depend on the depth of the exposures: a very shallow \ha\ image,
for instance, will only allow to identify stars with a strong emission line.
This effect is hard to quantify, as TTauri stars are variable and the
features we use will vary significantly at different times. In addition,
and more importantly, X-ray studies in the Milky Way \cite{walt86} showed that
Classical TTauri stars represent a minority of all low mass pre-Main Sequence
objects. Unfortunately the so-called Weak TTauri stars do not have any
clear photometric signature of their nature and they can be identified only
either in the X-rays or with spectroscopy.

\begin{figure}[!ht]
\begin{center}
\includegraphics[width=0.45\textwidth]{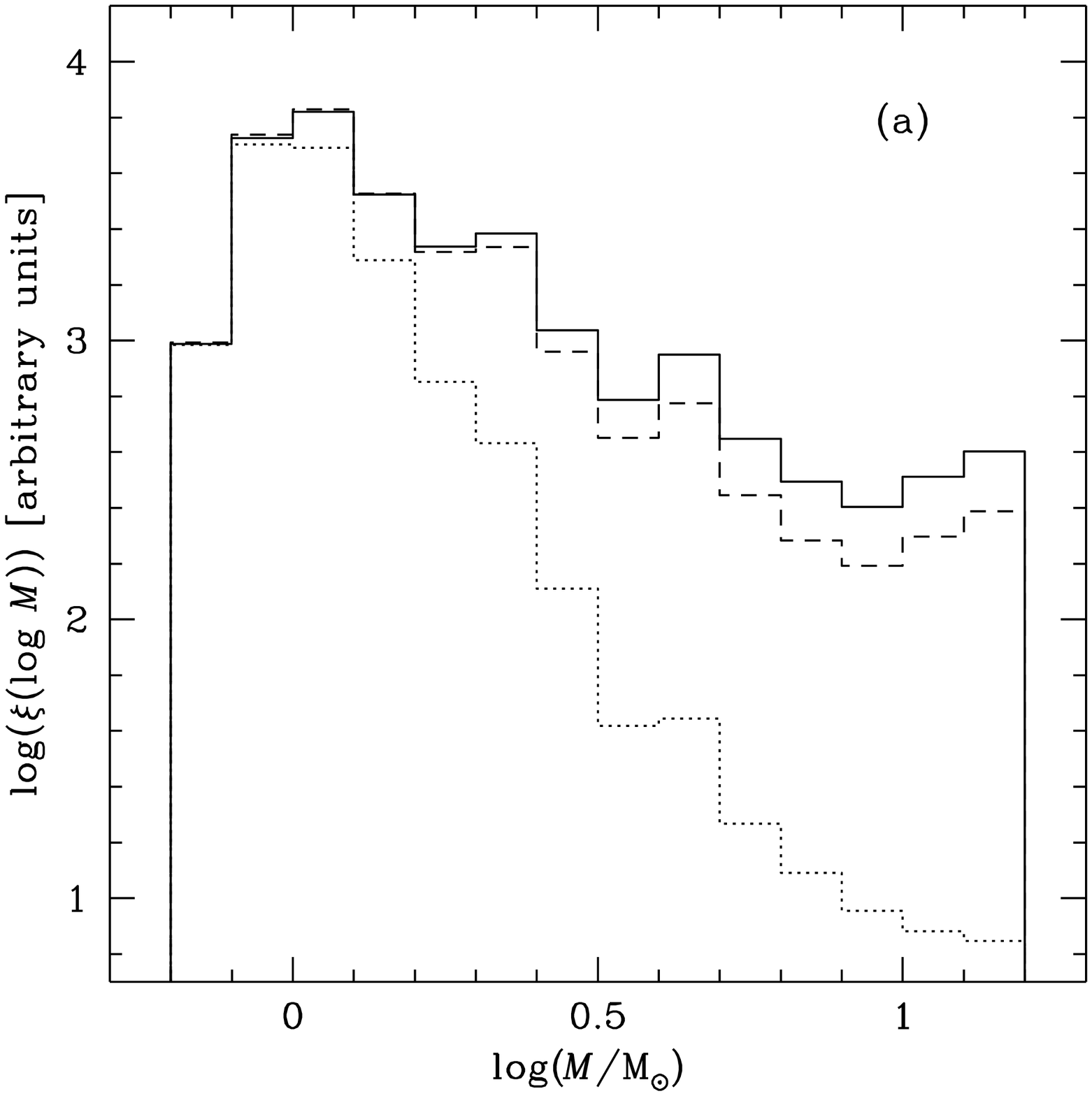}
\includegraphics[width=0.45\textwidth]{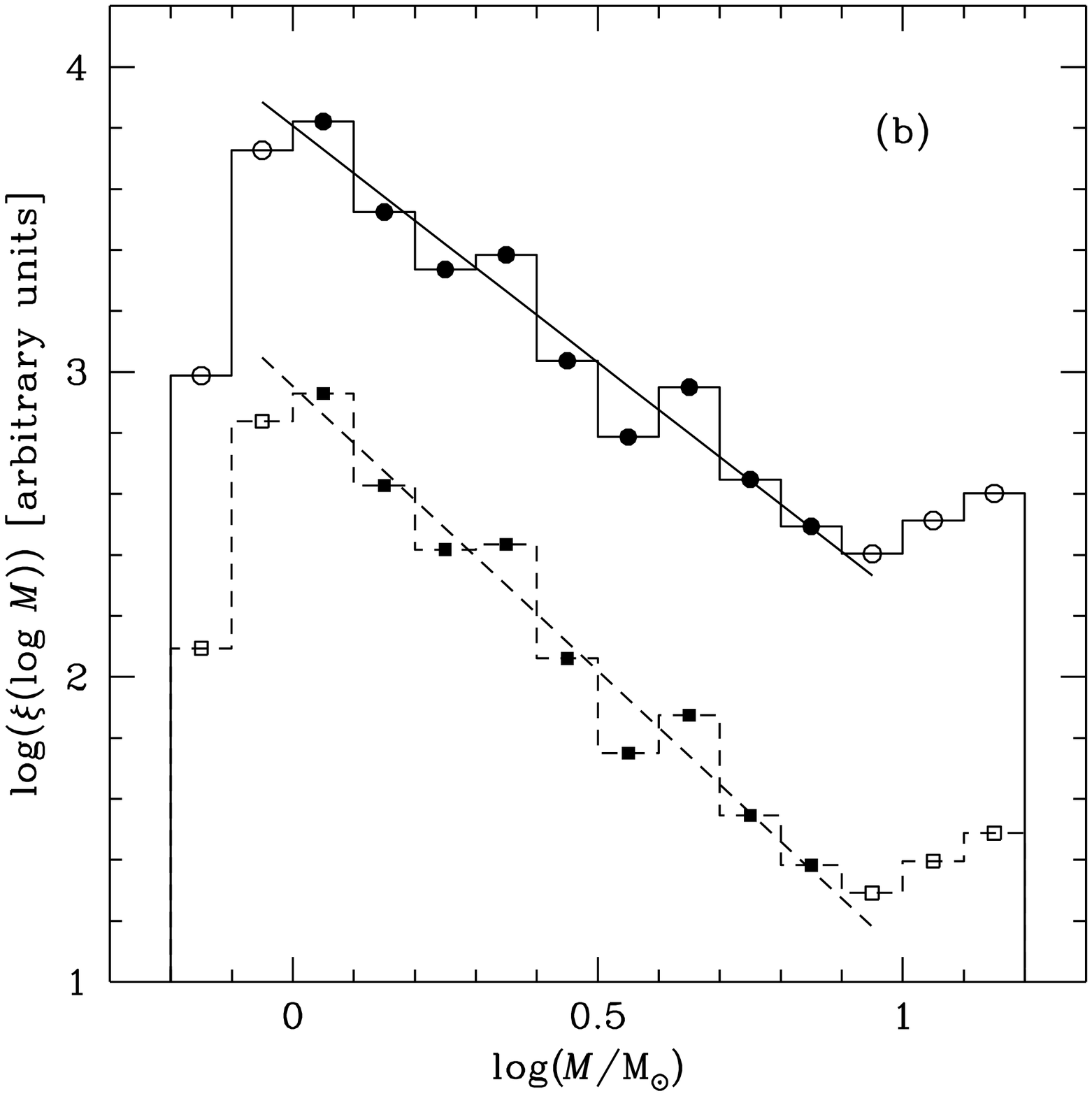}
\end{center}
\caption[]{Initial Mass Function in the neighborhood of SN~1987A.
\emph{Panel~(a):} the IMF derived including as TTauri stars only the stars
with \ha\ excess is shown as a full line, the one computed including also
the stars with U-band excess as a dashed line. The Present Day Mass Function
is also shown as a dotted line. \emph{Panel~(b):} power-law fit to the IMFs
of panel~(a). The bins used for the fit are marked with dots yielding a slope
of $\Gamma=-1.55$ if only the stars with \ha\ emission are included and
$\Gamma=-1.87$ if also the ones with U-band excess are considered. An
arbitrary shift is applied to better show the data.}
\label{fig:s87a_imf}
\end{figure}

An example of the effects of the incompleteness in identifying
TTauri stars is illustrated in Fig.~\ref{fig:s87a_imf}. There the Initial
Mass Function between 1 and 10~$M_\odot$ is plotted for the two recipes to
identify pre-Main Sequence stars described above: \ha\ emission or U band
excess. The derived slope is $\Gamma=-1.55$ in the first case (the classical
Salpeter value is $\Gamma=-1.35$) and as steep as $\Gamma=-1.87$ in the latter
one!

A complete discussion on the young population around SN~1987A can be found
in \cite{pan00}.

\subsection{The Double Cluster NGC~1850}
\begin{figure}[!ht]
\begin{center}
\includegraphics[width=0.7\textwidth]{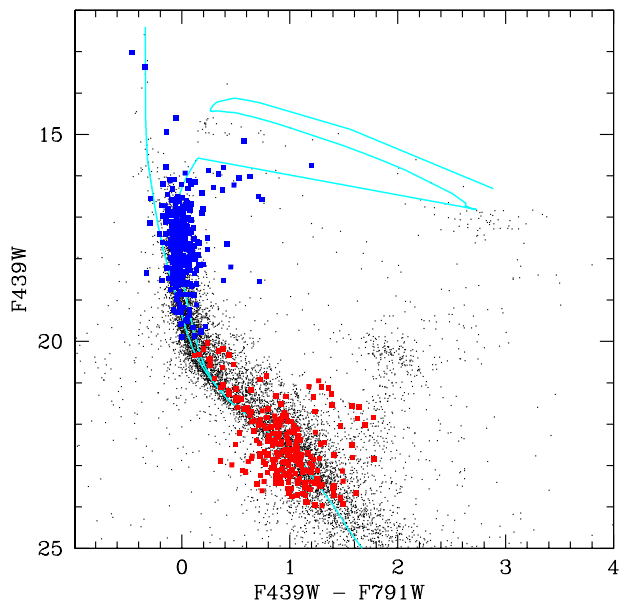}
\end{center}
\caption[]{F439W~vs~(F439W$-$F814W), \ie roughly B~vs~(B-I), color-magnitude
diagram of NGC~1850. The squares are \ha-emitting stars and the isochrones
for 4 and 50~Myr are displayed as full lines}
\label{fig:n1850_cmd}
\end{figure}

NGC~1850 is a double cluster in the outskirts of the LMC bar
(see Fig.~\ref{fig:lmc-dss}). According to our early WFPC2 investigation
\cite{gil94}, the main component, NGC1850A, has an age of $50\pm10$~Myr and
the slope of the IMF is $\alpha=-1.4\pm0.2$, \ie considerably flatter than the
Salpeter value of $-2.35$. NGC1850B, on the other hand, is extremely young,
$4\pm1$~Myr, and is characterized by a much steeper Initial Mass Function:
$\alpha=-2.6\pm0.1$. In addition, there are the usual LMC field stars, as
clearly indicated by the presence of the Red Clump at F439W$\simeq 20$,
F439W$-$F814W$\simeq 2.2$. The color-magnitude diagram for NGC~1850 is
shown in Fig.~\ref{fig:n1850_cmd}.

Once again, we know that there \emph{must} be TTauri stars associated
with the young cluster, but, as before, they are drowned in the much older
sub-giant population and broad band photometry only allows for statistical
arguments, but not an identification on a star to star basis. However, the
addition of \ha\ photometry, again with WFPC2, allowed us to discover 230
Classical TTauri candidates (and 350 Be stars belonging to the older
cluster). Let us stress again that the sample is by far incomplete and this
number surely is a lower limit to the real content of low mass pre-Main
Sequence stars. Once again, follow-up spectroscopy is needed to shed light
on the low mass star population.

The full analysis of NGC~1850 will appear in \cite{romprep}.

\section{VLT-FORS1 Spectroscopy}
To recapitulate, in order to fully characterize
young stellar populations projected on old field stars one has to find
a way to distinguish TTauri stars from field subgiants. Both methods we
have used, U-band excess and \ha\ emission, have allowed us to identify
several hundred TTauri candidates in the two regions we have
targeted. However, neither criterion yields a complete census of
low mass pre-Main Sequence stars. In particular, only Classical TTauri
stars can be identified, while \emph{all} Weak TTauri stars will be missed.

Ideally, a suitably deep X-ray survey would provide a complete sample, but,
unfortunately, the current generation of X-ray instruments does not have
enough sensitivity to detect TTauri stars beyond the Milky Way in a
reasonable integration time. In this case, then, even the Large Magellanic
Cloud is too distant! Thus, in order to understand the biases introduced by
the selection criteria we had to adopt on the WFPC2 imaging data, we
have applied for, and were granted, two Visitor Mode nights with FORS1 on the
VLT Antu (UT1) telescope in its Multi Object Spectroscopy
mode. The grism we have chosen, GRIS~300V, covers a wide spectral region
centered roughly at 5000~\AA\ and including H$\beta$, \ha\ and
Li\,{\sc i}~6707\,\AA. The sample selected for follow-up spectroscopy
consists of 20 candidate TTauri stars and as many stars that fall in the
same region of the HR diagram, but without neither \ha\ emission nor U-band
excess.

The observations were designed to fulfill two main goals. First, the
spectra would provide a critical test of our selection criterion
based on \ha\ emission and, second, they would allow to accurately determine
the characteristics of our putative TTauri stars (spectral type, amount of
veiling, line profile and equivalent width of the Balmer lines, etc).

Regrettably, though, we were not able to fulfill any of the proposed goals. The
unfortunate combination of \emph{El Ni\~no} and the Bolivian winter at the
beginning of the year 2000 resulted in our two nights having a seeing
variable between 1.5 and 2\arcsec: way too much for spectroscopy of
$V=21.5-22$ objects in a crowded field, even with an 8-meter telescope!!
As a partial consolation, among other
things, in those nights we did obtain several narrow band images of the
region of SN~1987A and of NGC~1850, which we have used to complete our
understanding of them by studying, in addition to the stars, also the
interstellar medium entwined with them. As for our original goal, at the time
of writing we have resubmitted the proposal to the ESO OPC for Period~68.
This time, though, in Service Mode\dots

\end{document}